\documentclass[twocolumn,nofootinbib,longbibliography,prx,superscriptaddress]{revtex4-1}

\usepackage{feynmp}
\usepackage{amsmath}
\usepackage{graphicx}
\usepackage{multirow}
\usepackage{latexsym}
\usepackage{textcomp}
\usepackage{verbatim}
\usepackage{color}
\usepackage{bm}
\usepackage{subfigure}
\usepackage{everysel}
\usepackage{keyval}
\usepackage{ragged2e}
\usepackage{dsfont}
\usepackage{amssymb}
\usepackage{enumitem}
\usepackage{pstricks}
\usepackage{mathtools}
\usepackage{braket}
\usepackage{slashed} 
\usepackage[colorlinks=true]{hyperref}

\usepackage{xr}

\usepackage[vcentermath]{youngtab}
\usepackage{ytableau}
\usepackage{float}
\usepackage{tabularx}

\usepackage{MnSymbol}

\usepackage{graphicx}
\usepackage{xcolor}
\usepackage{amsmath}
\usepackage{amsfonts}
\usepackage{bbm}

\usepackage{ulem}

\newcommand{\osum}{{%
    \setbox0\hbox{\circ}%
    \rlap{\hbox to \wd0{\hss\sum\hss}}\box0
}}

\newcommand{\tr}{{\textrm{tr}}}

\newcommand{\cA}{{\mathcal{A}}}

\newcommand{\cH}{{\mathcal{H}}}

\newcommand{\cO}{{\mathcal{O}}}

\newcommand{\cN}{{\mathcal{N}}}

\newcommand{\sun}{\text{\tiny SU(N)}}
\newcommand{\un}{\text{\tiny U(N)}}

\newcommand{\lyng}{\text{\tiny L}}
\newcommand{\ryng}{\text{\tiny R}}

\newcommand{\trun}{\text{\tiny trunc}}

\begin{document}

\title{Emergent factorization of Hilbert space at large $N$ and black hole}

\author{Antal Jevicki}
\thanks{Electronic Address: antal\_jevicki@brown.edu}
\affiliation{Department of Physics, Brown University, 182 Hope Street, Providence, RI 02912, USA}

\author{Debangshu Mukherjee}
\thanks{Electronic Address: debangshu.mukherjee@apctp.org}
\affiliation{Asia Pacific Center for Theoretical Physics, POSTECH, 77 Cheongam-ro, Nam-gu, Pohang-si, Gyeongsangbuk-do, 37673, Korea}

\author{Junggi Yoon}
\thanks{Electronic Address: junggi.yoon@apctp.org}
\affiliation{Asia Pacific Center for Theoretical Physics, POSTECH, 77 Cheongam-ro, Nam-gu, Pohang-si, Gyeongsangbuk-do, 37673, Korea}
\affiliation{Department of Physics, POSTECH, 77 Cheongam-ro, Nam-gu, Pohang-si, Gyeongsangbuk-do, 37673, Korea}
\affiliation{School of Physics, Korea Institute for Advanced Study, 85 Hoegiro Dongdaemun-gu, Seoul 02455, Korea}

\date{\today}

\begin{abstract} 
We investigate the emergent factorization of Hilbert space in the low-energy description of matrix models, addressing key aspects of the black hole information paradox. We examine the collective description for the low-energy sector of $SU(N)$ matrix model, characterized by a factorized Hilbert space composed of a finite number of boxes and anti-boxes. This factorization leads us to examine the emergence of thermofield dynamics~(TFD) state in the low energy sector from a fine-tuned state. In addition, we study the collective Hamiltonian of the $U(N)$ matrix model for the semi-classical description of ``particle-hole'' fluctuations around a background Young tableau. Our investigation of these matrix models elucidates a concrete mechanism for constructing the truncated algebra of accessible observables, thereby facilitating an understanding of black hole complementarity. In the context of the black hole information paradox, we discuss the origin of the island appearing inside the black hole and provide a reinterpretation of the recent proposal---the holography of information.

\end{abstract}

\maketitle

\section{Introduction}
\label{sec: introduction}

The study of emergent collective phenomenon in physics has unveiled profound insights into complex systems, where individual components interact to manifest new and often unexpected collective behaviors, including superconductivity, magnetism and solitons. From the perspective of the AdS/CFT correspondence~\cite{Maldacena:1997re}, semi-classical gravity can also be viewed as an emergent collective phenomenon in the large $N$ CFT~\cite{Das:1990kaa,deMelloKoch:2010wdf,Jevicki:2011ss,Das:2012dt,deMelloKoch:2014vnt,deMelloKoch:2014mos,Jevicki:2015sla,Jevicki:2016bwu,deMelloKoch:2018ivk}, emphasizing the emergence of the holographic coordinate from the boundary CFT.

Black holes stand as another quintessential example of complex systems in Physics, epitomizing the intricacies of emergent collective phenomena in gravity. The study of black holes, particularly in the context of the black hole information paradox~\cite{Hawking:1975vcx,Hawking:1976ra}, poses one of the most challenging puzzles in theoretical physics. The essence of the paradox lies in the entanglement between Hawking radiation and the inside of the black hole. Consequently, understanding the nature of operators inside black holes, known as \textit{mirror operators}~\cite{Papadodimas:2012aq,Papadodimas:2013jku}, is essential for the black hole information paradox.

Recently, black hole microstates have been investigated through cohomology for BPS operators in $\cN=4$ supersymmetric Yang-Mills theory~(SYM)
~\cite{Chang:2022mjp,Choi:2023znd,Choi:2023vdm,Chang:2024zqi}, where BPS \text{black hole operator}, defined by non-graviton cohomologies, have been explicitly identified for small values of $N=2,3,4$. This explicit construction of the black hole operators raises a question regarding the black hole information paradox: What is the microscopic origin of the mirror operators in the framework of constructing the BPS black hole operator? If black hole operators are formulated within a single $\cN=4$ SYM theory in large $N$ limit, the origin of operators inside the black hole could also be traced back to that SYM theory. This aligns with black hole complementarity~\cite{tHooft:1990fkf,Susskind:1993if,Papadodimas:2012aq,Papadodimas:2013jku,Papadodimas:2013wnh,Papadodimas:2015jra,Banerjee:2016mhh,deBoer:2018ibj,DeBoer:2019yoe}, wherein a simple operator inside the black hole can be represented as a complicated operator on the outside.

In this work, we study a simple matrix model as a toy model to demonstrate that in large $N$, the mirror operator in the black hole emerges as a collective excitation, analogous to a hole in the band theory. We show how the low-energy description for the matrix model gives rise to an emergent factorization of the Hilbert space, providing the underlying mechanism for black hole complementarity.

In the operator algebra of the BPS operators, the product of two BPS black hole operators can yield another BPS operator corresponding to larger BPS black hole. This differs from the operator algebra of the quantum field theory on the curved background, where the background does not take a part in the operator product. To understand the role of the background in the emergent collective excitation, we examine the semi-classical description for collective excitations around a large Young tableau which is analogous to the giant graviton \cite{McGreevy:2000cw,Grisaru:2000zn,Hashimoto:2000zp,Das:2000fu,Das:2000st,Corley:2001zk,Balasubramanian:2001nh} or the LLM geometry~\cite{Lin:2004nb,deMelloKoch:2008hen}. Our discussion extends to the emergent factorized Hilbert space in the semi-classical description around the background and its breakdown at finite $N$, thereby highlighting the constraints of the matrix model as a cutoff in the accessible observables.

Based on the emergent factorization of the Hilbert space, we elaborate old and new progresses in the black hole information paradox. We demonstrate that in the low energy sector, the Thermofield dynamics~(TFD) state can emerge from a fine-tuned state at finite $N$, shedding light on the factorization problem in the eternal black hole~\cite{Jevicki:2015sla,Harlow:2015lma,Guica:2015zpf,Harlow:2018tqv}. Furthermore, in the context of the entanglement entropy of Hawking radiation, we elucidate the origin of the \textit{island}~\cite{Almheiri:2019psf,Almheiri:2019hni,Almheiri:2019yqk,Penington:2019kki,Almheiri:2019qdq,Almheiri:2020cfm,Bak:2020enw} that appears inside the black hole and provide a reinterpretation of the \textit{holography of information}~\cite{Laddha:2020kvp,Chowdhury:2020hse,Raju:2020smc,Chowdhury:2021nxw,Raju:2021lwh,Chakraborty:2023los,Chakravarty:2023cll,deMelloKoch:2022sul} from the perspective of emergent factorization.

The paper is organized as follows. In Section~\ref{sec: su(n) case}, we investigate the collective description for the singlet sector of $SU(N)$ matrix model. We discuss the emergent factorization of Hilbert space in the low energy sector and its implications for understanding the mirror operator in the black hole. Section~\ref{sec: u(n) background case} expands on this discussion within the context of the $U(N)$ matrix model, focusing on the semi-classical description around the background Young tableau. Section~\ref{sec: black hole} is devoted to the implication of the emergent factorization for understanding the black hole information paradox. In Section~\ref{sec: discussion}, we make concluding remarks with open problems.

\section{Emergent Factorized Hilbert Space in Low Energy Sector of $SU(N)$ Matrix model}
\label{sec: su(n) case}

We begin by the one-dimensional $SU(N)$ matrix model whose Hamiltonian is given by
\begin{align}
	H\,=\, \tr \bigg[ U {\partial \over \partial U }U {\partial \over \partial U }\bigg]\ .\label{eq:SUN-Hamiltonian}
\end{align}
where $U(t)$ is the $N\times N $ $SU(N)$ matrix. This Hamiltonian is invariant under the $U(N)$ transformation, $U\to V^{-1}UV$ where $V\in U(N)$. Hence the $U(N)$-singlet sector can be expressed in terms of $U(N)$-invariant collective field defined by
\begin{align}
	\phi_n\,\equiv \, \tr \big(U^n\big) \qquad (n=1,2,\cdots)\;\ .
\end{align}
The collective field $\phi_n$ is also called as a plaquette variable of winding number $n$ in the (one-plaquette) Kogut-Susskind lattice gauge theory~\cite{Kogut:1974ag,Jevicki:1980zq}. Note that not all $\phi_n$'s are independent at finite $N$. $\{\phi_1,\phi_2,\cdots, \phi_{N-1}\}$ forms the set of independent collective fields, and other collective field, $\phi_m$ $(m\geqq N)$, can be expressed in terms of them.

In large $N$ limit, the collective fields $\phi_n$ ($n\geqq 1$) become independent, and the dynamics of the $U(N)$ singlet sector can be described by the collective Hamiltonian of the collective fields~\cite{Jevicki:1980zg,Jevicki:1991yi}:
\begin{align}
	H\,=\,& N\sum_{n=1}^\infty  n \phi_n {\partial \over \partial\phi_n} +\sum_{n=2}^\infty \sum_{m=1}^{n-1} n \phi_m\phi_{n-m} {\partial \over \partial \phi_n}\cr
 &\hspace*{-0.5cm} + \sum_{n=1}^\infty \sum_{m=1}^\infty nm \phi_{n+m} {\partial\over \partial\phi_n}{\partial\over \partial\phi_m} - {1\over N}\bigg[\sum_{n=1}^\infty n\phi_n{\partial \over \partial \phi_n} \bigg]^2\ .\label{eq:SUN-collH}
\end{align}
Here, the last term of order $\cO(N^{-1})$ comes from the $SU(N)$ condition, $\det U=1$. The cubic interaction terms represents the splitting and joining of the loops, with the coefficients indicating the number of possible ways these processes can occur~\cite{Jevicki:1991yi,Avan:1995sp,Jevicki:2013kma,Jevicki:2015irq}. The Fock space representation~\cite{NOMURA1986289} can be achieved by substituting $\phi_n$ with $\sqrt{n} a^\dag_n$ and ${\partial\over \partial \phi_n}$ with ${1\over \sqrt{n}}a_n$, respectively~\cite{Jevicki:1991yi}.

It was shown that the collective Hamiltonian~\eqref{eq:SUN-collH} is diagonalized by the $SU(N)$ character represented by the Schur polynomials of the collective fields:
\begin{align}
	P(R;\{\phi\}) \,=\, {1\over |R|!}\sum_{g\in S_{|R|}} \text{ch}_R(g) \prod_{j=1}^{|R|} \big(\phi_j\big)^{\mu(g)_j}\ , \label{eq: character}
\end{align}
where the energy eigenstate is characterized by the Young tableau $R$, where $|R|$ denotes the total number of boxes in the Young tableau $R$. In addition, $\text{ch}_{R}(g)$ denotes the character of $S_{|R|}$ for the group element $g\in S_{|R|}$ in the irreducible representation associated to the Young tableau $R$. The element $g\in S_{|R|}$ is labelled by its cycle structure, $g=1^{\mu(g)_1}2^{\mu(g)_2}\cdots |R|^{\mu(g)_{|R|}}$ where $\mu(g)_j$ denotes the number of cycles of length $j$. The corresponding energy eigenvalue is given by the Casimir of the Young tableau $R$.
\begin{align}
E_{\sun}(R)\,=\, \sum_{j=1}^n r_j \big[N +r_j -(2j-1)\big]-\frac{1}{N}\bigg(\sum_{j=1}^n r_j\bigg)^2\ .\label{eq: casimir su(n)}
\end{align}
Here, $r_j$ denotes the number of boxes in the $j$th row in the Young tableau $R$.

\begin{figure}[t!]
\centering
\includegraphics[width=0.8\linewidth]{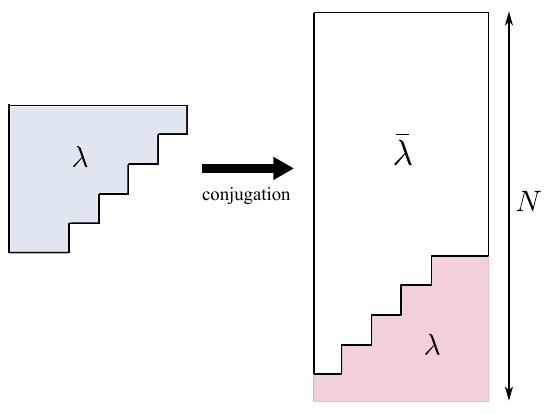}
	\caption{For a Young tableau $\lambda$, the conjugate Young tableau $\overline{\lambda}$ is defined by flipping $\lambda$ both vertically and horizontally. }
 \label{fig: conjugate young tableau}
\end{figure}

\begin{figure}[t]
\centering
\includegraphics[width=0.6\linewidth]{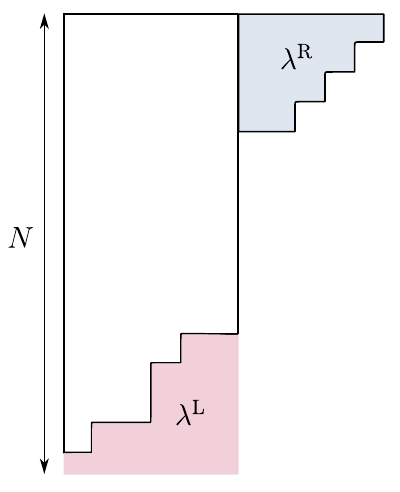}
	\caption{A generic Young tableau in the low energy sector of $SU(N)$ matrix model. It consists of $\cO(N^0)$ number of boxes~(blue) and anti-boxes (red), respectively.}
 \label{fig: generic low energy young tableau}
\end{figure}

The energy eigenvalue~\eqref{eq: casimir su(n)} of a Young tableau $R$ is identical to that of its conjugate Young tableau, $\overline{R}$, where the conjugate Young tableau $\overline{R}$ is defined by inverting $R$ both upside down and from left to right, and filling it out to a rectangle with $N$ rows (See Fig.~\ref{fig: conjugate young tableau}). For example the \textit{anti-box} ${\tiny \overline{\yng(1)}}$, a column of $N-1$ boxes, has the same energy as a single box ${\tiny \yng(1)}$, which is reminiscent of the particle-hole relationship. Therefore, in the low energy sector, a generic Young tableau consists of a Young tableau $\lambda^\lyng$ with a finite number of anti-boxes as well as a Young tableau $\lambda^\ryng$ with a finite number of boxes (See Fig.~\ref{fig: generic low energy young tableau}). The energy eigenvalue corresponding to such a Young tableau denoted by $R=(\lambda^\lyng, \lambda^\ryng)$ can be decomposed into that of $\lambda^\lyng$ and $\lambda^\ryng$ with $1/N$ correction.  
\begin{align}
	E_\sun(R)\,=\, E_\sun(\lambda^\lyng)+ E_\sun(\lambda^\ryng) + {2|\lambda^\lyng| \,| \lambda^\ryng |\over N}\ .\label{eq: casimir su(n) modified}
\end{align}
The details are given in Appendix~\ref{app: su(n) casimir}. From the point of view of the box, the low energy sector includes Young tableaux containing $\cO(N)$ numbers of boxes.

In order to describe the low energy sector of the $SU(N)$ matrix model~\eqref{eq:SUN-Hamiltonian}, the Hilbert space for the collective Hamiltonian~\eqref{eq:SUN-collH} needs to be expanded. The collective Hamiltonian in Eq.~\eqref{eq:SUN-collH} is formulated based on the collective field $\phi_n$ which are considered to be independent in large $N$ limit. However, this framework cannot accommodate Young tableaux with a number of boxes of order $\cO(N)$ due to the breakdown of the independence of the collective fields.

In the low energy sector, we define the two branches of the collective field as follows:
\begin{align}
	\phi_n\,\equiv\, \tr( U^n) \;\;,\quad \bar{\phi}_n\,=\, \tr (U^{-n})  \ .
\end{align}
While $\phi_n$ and $\bar{\phi}_m$ are not independent at finite $N$, they become independent in the large $N$ limit. In this low energy description, $\phi_n$ and $\bar{\phi}_n$ are unrestricted because we consider a finite number of boxes and anti-boxes in the large $N$ limit. In terms of $\phi_n$ and $\bar{\phi}_n$, the collective Hamiltonian, which reproduces the low energy spectrum~\eqref{eq: casimir su(n) modified}, is found to be
\begin{align}
	H\,=\,& N\sum_{n=1}^\infty  n \phi_n {\partial \over \partial\phi_n} +\sum_{n=2}^\infty \sum_{m=1}^{n-1} n \phi_m\phi_{n-m} {\partial \over \partial \phi_n} \cr 
 &+\sum_{n=1}^\infty \sum_{m=1}^\infty nm \phi_{n+m} {\partial\over \partial\phi_n}{\partial\over \partial\phi_m} -{1\over N}\bigg(\sum_{n=1}^N n\phi_n {\partial \over \partial\phi_n}\bigg)^2 \cr
	&+ N\sum_{n=1}^\infty  n \bar{\phi}_n {\partial \over \partial\bar{\phi}_n} +\sum_{n=2}^\infty \sum_{m=1}^{n-1} n \bar{\phi}_m\bar{\phi}_{n-m} {\partial \over \partial \bar{\phi}_n} \cr
 &+ \sum_{n=1}^\infty \sum_{m=1}^\infty nm \bar{\phi}_{n+m} {\partial\over \partial\bar{\phi}_n}{\partial\over \partial\bar{\phi}_m}- {1\over N}\bigg(\sum_{n=1}^\infty  n \bar{\phi}_n {\partial \over \partial\bar{\phi}_n}\bigg)^2 \cr
 &+ {2\over N} \sum_{n=1}^\infty  n \phi_n {\partial \over \partial\phi_n}\sum_{m=1}^\infty  m \bar{\phi}_m {\partial \over \partial \bar{\phi}_m}\ .
\end{align}

\begin{figure}[t!]
\centering
\includegraphics[width=\linewidth]{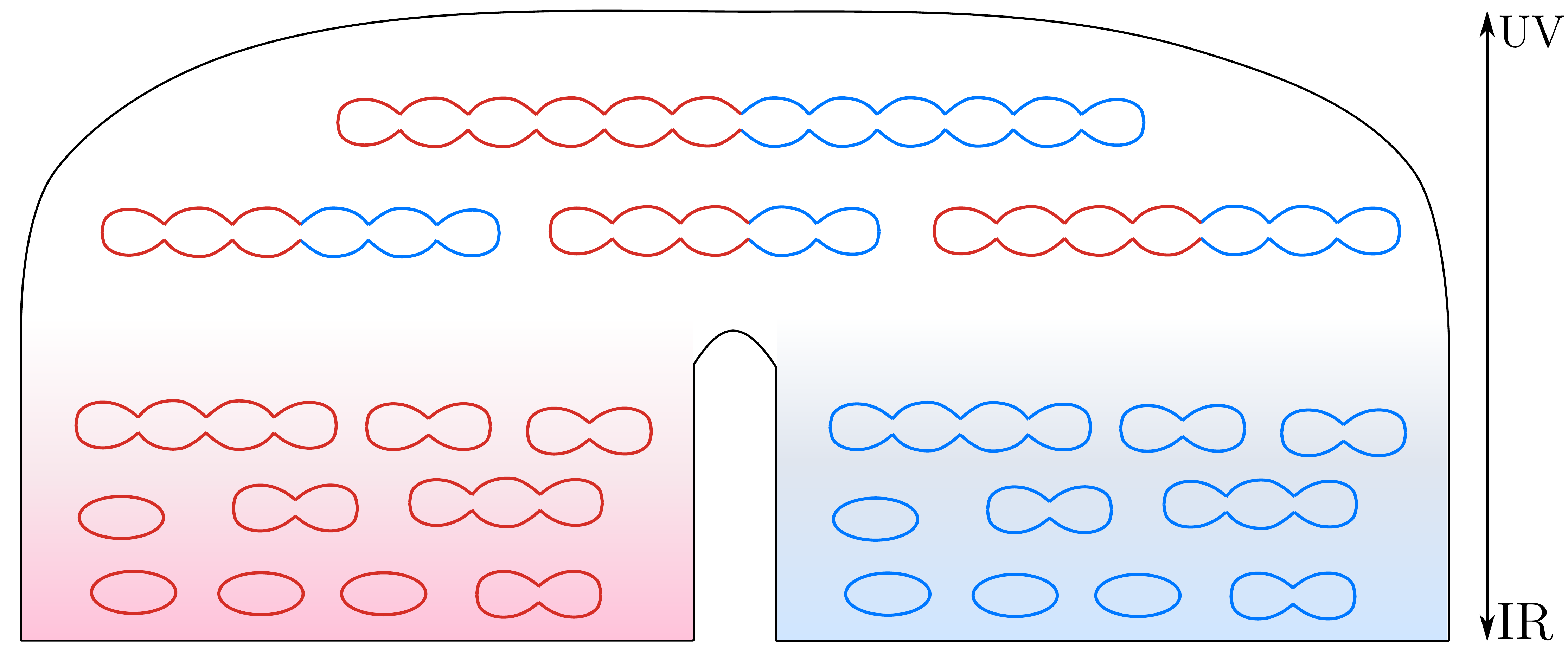}
	\caption{The emergent factorization of Hilbert space in the low energy sector. In the low energy description, the loops $\phi_n$ (blue) and anti-loops $\bar{\phi}_n$ (red) are independent. But, in the UV regime, they are no longer independent due to the constraints of the matrix model.}
 \label{fig: emergent factorization in low energy}
\end{figure}

The Hilbert space of the low energy sector is factorized into two Fock spaces, $\cH_\phi \otimes \cH_{\bar{\phi}}$, which is constructed by two branches of the collective fields, $\{\phi_n\}$ and $\{\bar{\phi}_n\}$, respectively. This factorization of the Hilbert space is an emergent phenomenon in the low energy sector because $\cH_{\bar{\phi}}$ is not independent of $\cH_\phi$ at finite $N$ (See Fig.~\ref{fig: emergent factorization in low energy}). For example, using the constraint in the Schur polynomial of $SU(N)$ matrix, the state corresponding to a single anti-box can be expressed in terms of that of the $N-1$ boxes at finite $N$ as follows (See Appendix~\ref{app: schur polynomial identity} for the detailed derivation).
\begin{align}
	&P({\tiny \yng(1)},\{\bar{\phi}\} )\,=\, P\big(\; \left. {\ytableausetup{smalltableaux,centertableaux} \tiny
	\begin{ytableau}
	 \\
	\none[\tiny \cdot]\\
	\none[\tiny \cdot]\\
	\none[\tiny \cdot]\\
	 \\
	\end{ytableau} }\,\right\}{\scriptstyle N-1} \; ,\{\phi\}\big)\ ,\cr
 	 \Longrightarrow \;\;&\bar{\phi}_1\,=\,{1\over (N-1)!}\phi_1^{N-1}+\cdots +{(-1)^{N}\over N-1}\phi_{N-1}\ .
\end{align}
Hence the collective description at finite $N$ should take into account this relation (See Ref.~\cite{deMelloKoch:2008hen}).

On the other hand, in the low energy sector, the constraints in the matrix model that dictates the relation between boxes and anti-boxes is absent, because we consider a finite number of boxes and anti-boxes. As a result, in the low energy sector the anti-boxes emerges as a independent degree of freedom, or in other words, a quasiparticle or a collective excitation of the box ${\tiny \yng(1)}$ similar to a hole in the band theory.

The energy eigenstate corresponding to a Young tableau $R=(\lambda^\lyng, \lambda^\ryng)$ is also factorized into two Schur polynomials of $\phi$'s and $\bar{\phi}$'s, respectively:
\begin{align}
	\Psi_{(\lambda^\lyng, \lambda^\ryng)}\,=\, P(\lambda^\lyng,\{\bar{\phi}\})P(\lambda^\ryng,\{\phi\}) \ .
\end{align}
Here, one should not take it simply as a product of two Schur polynomial. Since the Hilbert space in the low energy sector is factorized, it is a tensor product of two Schur polynomial. \textit{i.e.}
\begin{align}
	\Psi_{(\lambda^\lyng, \lambda^\ryng)}\,=\, \big(P(\lambda^\lyng,\{\bar{\phi}\}),P(\lambda^\ryng,\{\phi\}) \big)\ .
\end{align}
Therefore, one should not apply the fusion rule identity between the two Schur polynomials, $P(\lambda^\lyng,\{\bar{\phi}\})$ and $P(\lambda^\ryng,\{\phi\})$, as it is based on the constraint of the matrix absent in the low energy description.
%
%
%
%
%
%

\section{Emergent Factorized Hilbert Space in Semi-classical Description of $U(N)$ Matrix model}
\label{sec: u(n) background case}

We now consider the semi-classical description for the $U(N)$ matrix model of which the Hamiltonian is of the same form as in Eq.~\eqref{eq:SUN-Hamiltonian}. The energy eigenstate of the $U(N)$ matrix model is also written as the $U(N)$ character associated with the Young tableau $R$ in~Eq.~\eqref{eq: character}. The corresponding energy eigenvalue is given by
\begin{align}
E_{\un}(R)\,=\, \sum_{j=1}^n r_j \big[N +r_j -(2j-1)\big]\ .\label{eq: casimir u(n)}
\end{align}
In the $U(N)$ matrix model, the energy associated with a single box, ${\tiny \yng(1)}$, differs from that of a single anti-box, ${\tiny \overline{\yng(1)}}$ in contrast to $SU(N)$ case. Furthermore, in the $U(N)$ case, columns that are completely filled with boxes are not trivial. Therefore  a reference Young tableau is required to define a new type of conjugation, which gives rise to particle-hole relationship in the low energy sector.

\begin{figure}[t!]
\centering
\includegraphics[width=\linewidth]{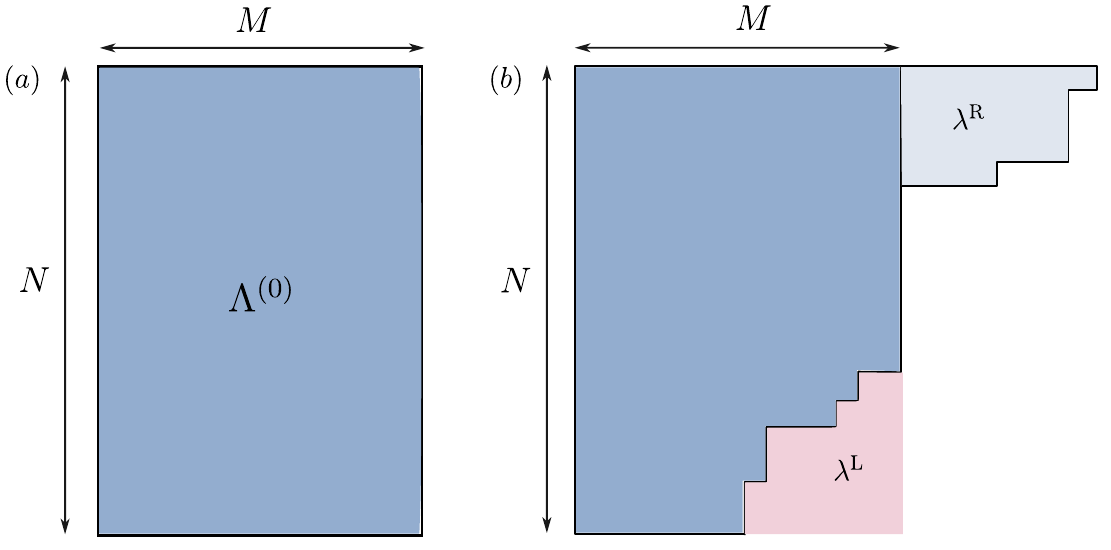}
	\caption{(a) Background $N\times M$ Young tableau $\Lambda^{(0)}$.  (b) A generic Young tableau in the semi-classical description of $U(N)$ matrix model. It is composed of $\cO(N^0)$ number of boxes (light blue) and anti-boxes (red) together with background (dark blue)}
 \label{fig: background young tableau}
\end{figure}

\begin{figure}[t!]
\centering
\includegraphics[width=0.8\linewidth]{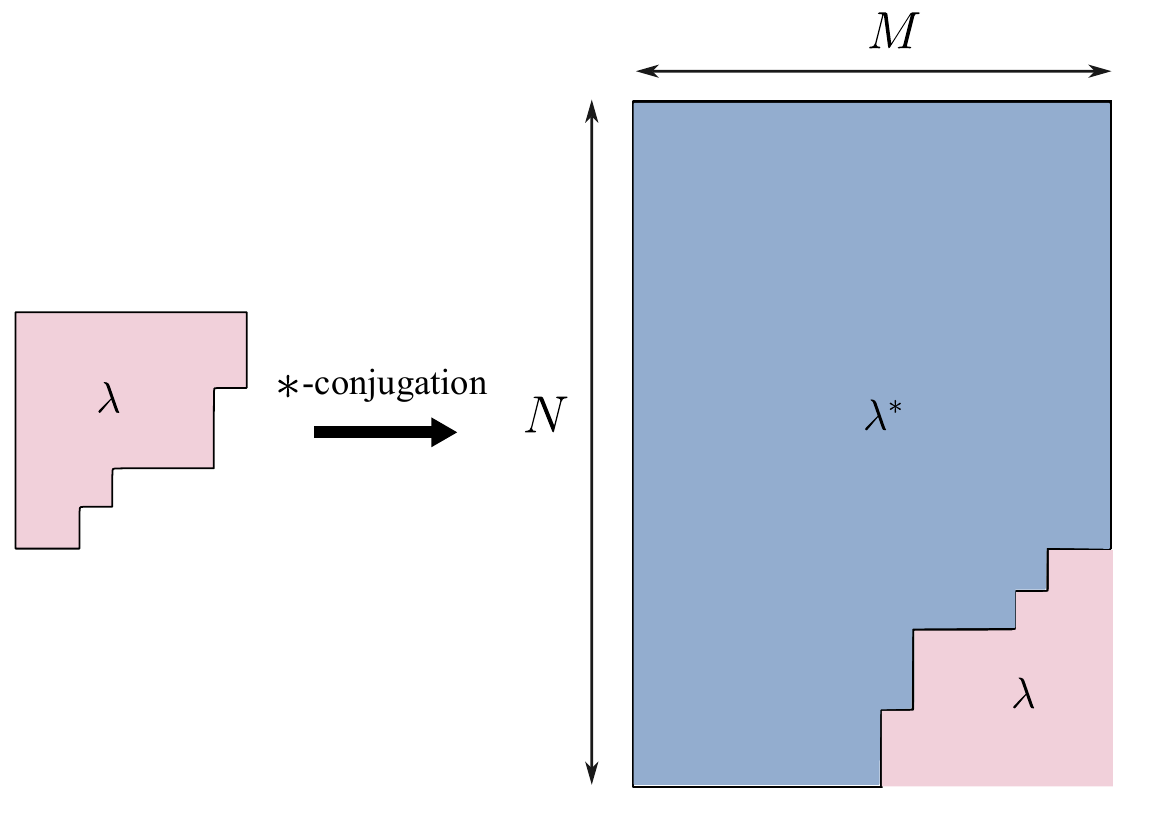}
	\caption{$\ast$-conjugation of Young tableau $\lambda$ with respect to background Young tableau $\Lambda^{(0)}$. }
 \label{fig: star conjugation}
\end{figure}

For the semi-classical description of the $U(N)$ matrix model, we introduce a background $N\times M $ Young tableau, $\Lambda^{(0)}$ ($N,M\gg 1$), which plays a similar role of a giant graviton \cite{McGreevy:2000cw,Grisaru:2000zn,Hashimoto:2000zp,Das:2000fu,Das:2000st,Corley:2001zk,Balasubramanian:2001nh} or a black hole background in the gravity (See Fig.~\ref{fig: background young tableau} (a)).
\begin{align}
    \Lambda^{(0)}_j\,\equiv\,M \qquad (j=1,2,\cdots, N)\ .
\end{align}
Furthermore, we define a $\ast$-conjugation of a Young tableau $\lambda$ with respect to the background $\Lambda^{(0)}$ by subtracting $\lambda$ from $\Lambda^{(0)}$ (See Fig.~\ref{fig: star conjugation}).
\begin{align}
\label{eq:starconjugation}
    (\lambda^\ast)_j\,\equiv\,M-\lambda_{N+1-j}\qquad (j=1, 2,\cdots, N)\ .
\end{align}
This $\ast$-conjugation is well-defined for Young tableaux that do not exceed the background $\Lambda^{(0)}$, which will be sufficient for the semi-classical description.

In our semi-classical description, we consider fluctuations around the background, $\Lambda^{(0)}$, which form the Hilbert space. A generic Young tableau in the Hilbert space, denoted by $(\lambda^\lyng,\lambda^\ryng)$, is characterized by the configuration where the Young tableau $\lambda^\ryng$ with a finite number boxes is attached to the background Young tableau $\Lambda^{(0)}$ and the Young tableau $\lambda^\lyng$ with a finite number boxes is subtracted from $\Lambda^{(0)}$ (See Fig.~\ref{fig: background young tableau} (b)). The Young tableau $\lambda^\lyng$ is a collective excitation with respect to the background, and $\lambda^\ryng$ and $\lambda^\lyng$ play a role of particle and hole, respectively.

The Hilbert space in the semi-classical description around the background is factorized into two Fock spaces, $\cH_\phi \otimes \cH_{\bar{\phi}}$, generated by the collective fields, $\phi_n$ and $\bar{\phi}_n$:
\begin{align}
    \phi_n\,\equiv\, \tr(U^n)\;,\quad \bar{\phi}_n\,\equiv\,\tr(U^{-n})\qquad (n=1,2,3,\cdots)\ .
\end{align}
The emergent factorization of the Hilbert space in the semi-classical description will hold as long as we consider a finite number of boxes for $\lambda^\lyng$ and $\lambda^\ryng$. In this regime, the energy eigenvalue of a state associated with a Young tableau $R=(\lambda^\lyng,\lambda^\ryng)$ is found to be 
\begin{align}
	&E_\un(R)\cr
 \,=\,&
 NM^2 + E_\un(\lambda^\ryng) + E_\un(\lambda^\lyng)+2M\big( \big|\lambda^\ryng\big| - \big| \lambda^\lyng \big|  \big)\ ,\cr
 \,=\,& N\big(M^2 +\big|\lambda^\ryng\big| + \big| \lambda^\lyng \big|  \big)  + 2M\big( \big|\lambda^\ryng\big| - \big| \lambda^\lyng \big|  \big) \cr
 &+ \sum_{j} \big[ \lambda^\ryng_j (\lambda^\ryng_j+1-2j)+  \lambda^\lyng_j(\lambda^\lyng_j  +1-2j)\big]\ , \label{eq: u(n) energy semi classical}
\end{align}
where we show the derivation in Appendix~\ref{app: u(n) casimir background}. In the limit $N\gg M\gg 1$, this can be indeed viewed as a fluctuation around the background energy $NM^2$.

In the semi-classical regime, the collective Hamiltonian that dictates the dynamics of the fluctuations is determined to be
\begin{align}
	H\,=\,&NM^2 +M \sum_{n=1}^\infty  n \bigg(\phi_n {\partial \over \partial\phi_n}-\bar{\phi}_n {\partial \over \partial\bar{\phi}_n}\bigg)  \cr
	&+ N\sum_{n=1}^\infty  n \phi_n {\partial \over \partial\phi_n} +\sum_{n=2}^\infty \sum_{m=1}^{n-1} n \phi_m\phi_{n-m} {\partial \over \partial \phi_n}\cr
 &+ \sum_{n=1}^\infty \sum_{m=1}^\infty nm \phi_{n+m} {\partial\over \partial\phi_n}{\partial\over \partial\phi_m}  \cr
	&+ N\sum_{n=1}^\infty  n \bar{\phi}_n {\partial \over \partial\bar{\phi}_n} +\sum_{n=2}^\infty \sum_{m=1}^{n-1} n \bar{\phi}_m\bar{\phi}_{n-m} {\partial \over \partial \bar{\phi}_n} \cr
 &+ \sum_{n=1}^\infty \sum_{m=1}^\infty nm \bar{\phi}_{n+m} {\partial\over \partial\bar{\phi}_n}{\partial\over \partial\bar{\phi}_m}\ .
\end{align}
The energy eigenstate corresponding to a Young tableau $R=(\lambda^\lyng, \lambda^\ryng)$ is expressed as
\begin{align}
    \Psi_{(\lambda^\lyng, \lambda^\ryng)}\,=\, P (\lambda^\lyng,\{\bar{\phi}\})\, P (\lambda^\ryng,\{\phi\})\ , \label{eq: u(n) state}
\end{align}
with its eigenvalue yielding the semi-classical spectrum in Eq.~\eqref{eq: u(n) energy semi classical}. As in the $SU(N)$ case, the state in Eq.~\eqref{eq: u(n) state} should be interpreted as a direct product rather than a simple product.

In addition, the factor $[\det U]^M$, representing the background, is stripped off in the realization of the state~\eqref{eq: u(n) state} for the fluctuations. At finite $N$, the Schur polynomial for the background Young tableau $\Lambda^{(0)}$ is expressed in terms of the determinant operator:
\begin{align}
    P(\Lambda^{(0)},U) \,=\, \big[\det U\big]^M\ .
\end{align}
However, the determinant operator $\det U$ does not belong to the Hilbert space $\cH_\phi\otimes \cH_{\bar{\phi}}$ since we take into account $\cO(N^0)$ numbers of the boxes and anti-boxes for the semi-classical description. Hence the determinant, $\det U$, should be regarded as a $c$-number instead of an operator acting on the Hilbert space.

In the semi-classical description with the realization of the fluctuating state in Eq.~\eqref{eq: u(n) state}, the operator product should be evaluated among $\phi$'s and among $\bar{\phi}$'s respectively, not between $\phi$ and $\bar{\phi}$.
\begin{align}
    (\lambda_1,\mu_1)\times(\lambda_2,\mu_2)\,=\, \sum_{\lambda_3,\mu_3}{c_{\lambda_1\lambda_2}}^{\lambda_3}{c_{\mu_1\mu_2}}^{\mu_3}(\lambda_3,\mu_3)\ ,
\end{align}
where ${c_{\lambda \mu}}^\nu$ is the tensor product coefficient of $U(N)$ group. The operator product between $\phi$ and $\bar{\phi}$ is involved with the constraint of the matrix model such as
\begin{align}
    P\big(\; \left. {\ytableausetup{smalltableaux,centertableaux} \tiny
	\begin{ytableau}
	 \\
	\none[\tiny \cdot]\\
	\none[\tiny \cdot]\\
	\none[\tiny \cdot]\\
	 \\
	\end{ytableau} }\,\right\}{\scriptstyle N} \; ,\{\phi\}\big)\,=\,&{1\over N!}\phi_1^{N}+\cdots +{(-1)^{N+1}\over N}\phi_{N}\ ,\cr
    \,=\,& \det U \ ,
\end{align}
which is absent in the semi-classical description. In addition, when realizing the fluctuating state as described in Eq.~\eqref{eq: u(n) state}, the background is intact under the operator product. For example, in the operator product 
\begin{align}
    \Psi_{(\;\;\cdot\;\;,\;{\tiny \yng(1)}\;)}\times \Psi_{(\;\;\cdot\;\;,\;{\tiny \yng(1)}\;)}\,=\, \Psi_{(\;\;\cdot\;\;,\;{\tiny \yng(2)}\;)} + \Psi_{(\;\;\cdot\;\;,\;{\tiny \yng(1,1)}\;)} \ ,\label{eq: operator product example}
\end{align}
all $\Psi$'s in the left-hand and right-hand sides represent fluctuations with respect to same background $\Lambda^{(0)}$. This is analogous to the operator product algebra in the curved background~\cite{Leutheusser:2021qhd,Leutheusser:2021frk,Witten:2021jzq} where the operator product is analyzed semi-classically in the fixed background geometry. 

\begin{figure}[t!]
\centering
\includegraphics[width=\linewidth]{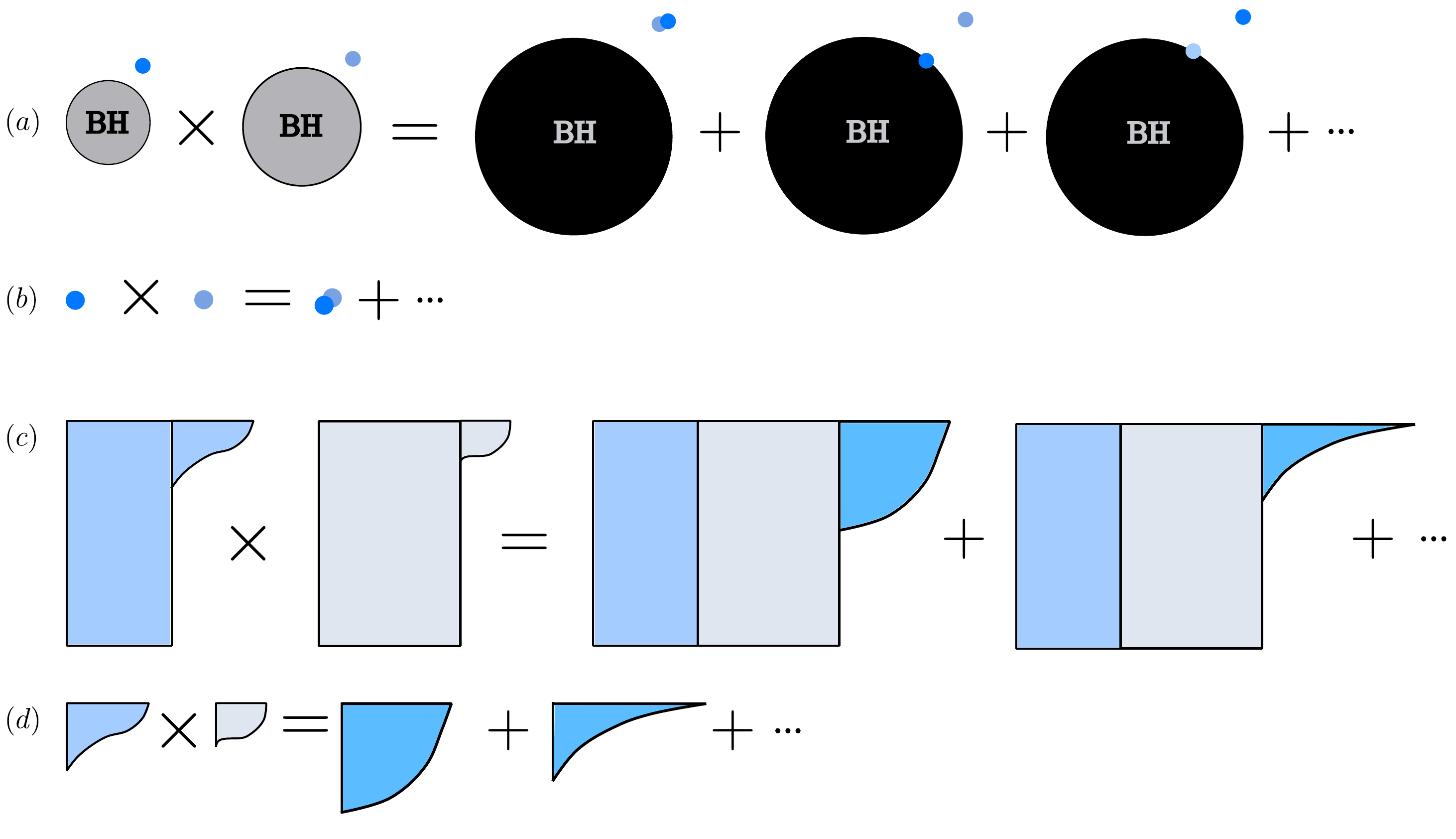}
	\caption{(a) The product of two BPS black hole operators, representing a black hole with graviton hair, can produce another BPS black hole operators with large black hole with graviton hair. It is necessary to take into account not only the OPE of the gravitons but also the OPE between the graviton and the black hole, \textit{etc}. (b) In the semi-classical description, the OPE is involved with graviton while the black hole background does not take part in. (c) The tensor product of two Young tableaux with background yields larger background analogous to the OPE of BPS black hole operators. (d) In the semi-classical description, the background Young tableau does not participate in the tensor product.}
 \label{fig: black hole operator product}
\end{figure}

On the other hands, the background Young tableau $\Lambda^{(0)}$ takes part in the operator product at finite $N$. This corresponds to the full quantum gravity calculations, in which the background geometry is also treated as an operator. For example, at finite $N$ the product of two operators, which consists of one column corresponding to the background ($M=1$) and a single box (and anti-box, respectively) representing the fluctuation around the background, is given by
\begin{align}
    {\ytableausetup{smalltableaux,centertableaux} \tiny
	\begin{ytableau}
	~ &  \\
    \\
	\none[\tiny \cdot]\\
	\none[\tiny \cdot]\\
	\none[\tiny \cdot]\\
	\\
    \\
	\end{ytableau} }\;\;\times \;\;{\ytableausetup{smalltableaux,centertableaux} \tiny
	\begin{ytableau}
	~  \\
    \\
	\none[\tiny \cdot]\\
	\none[\tiny \cdot]\\
	\none[\tiny \cdot]\\
	 \\
    \none   \\
	\end{ytableau} }\quad=\quad {\ytableausetup{smalltableaux,centertableaux} \tiny
	\begin{ytableau}
	~ &  &  \\
    ~ &  \\
	\none[\tiny \cdot] & \none[\tiny \cdot]\\
	\none[\tiny \cdot] & \none[\tiny \cdot] \\
	\none[\tiny \cdot] & \none[\tiny \cdot]\\
	~ &  \\
 ~  \\
	\end{ytableau} }\quad+\quad {\ytableausetup{smalltableaux,centertableaux} \tiny
	\begin{ytableau}
	~ &  \\
    ~ &   \\
	\none[\tiny \cdot] & \none[\tiny \cdot]\\
	\none[\tiny \cdot] & \none[\tiny \cdot]\\
	\none[\tiny \cdot] & \none[\tiny \cdot]\\
	~ &  \\
     ~ &  \\
	\end{ytableau} } \ .
\end{align}
Here, the first term on the right-hand side represents the separate product of the two fluctuations while two backgrounds (each one column) combine to create a larger background. The second term indicates the intermingling of $\phi$ and $\bar{\phi}$. In Fig.~\ref{fig: black hole operator product}, we depict an analogous situation in the context of BPS black hole operator~\cite{Chang:2022mjp,Choi:2023znd,Choi:2023vdm,Chang:2024zqi} in $\cN=4$ Yang-Mills theory.

\section{Implications for Black Holes}
\label{sec: black hole}

\subsection{Emergent TFD State}
\label{sec: emergent tfd state}

 \begin{figure}[t!]
\centering
\includegraphics[width=0.95\linewidth]{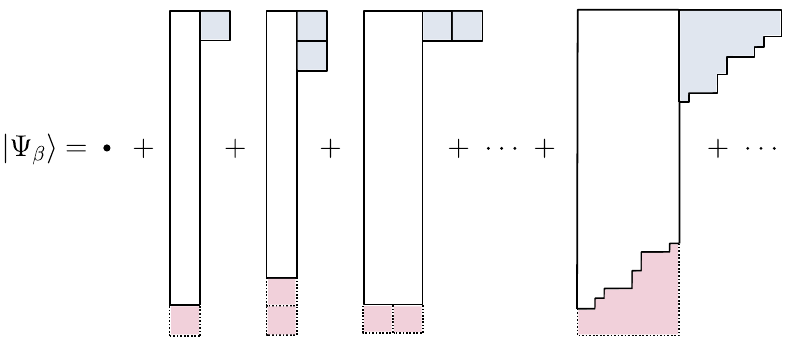}
	\caption{A fine-tuned state at finite $N$ which becomes Thermofield dynamics~(TFD) state in the low energy sector. It is a linear combination of the Schur polynomials with self-conjugate Young tableaux $R=(\lambda,\lambda)$. For a large Young tableau $\lambda$, the boxes (blue) and the anti-boxes (red) could come into contact, making the Young tableau $R=(\lambda,\lambda)$ ill-defined. Therefore, we will exclude those Young tableaux $\lambda$ that are also irrelevant in the low energy sector. }
 \label{fig: emergent tfd state}
\end{figure}

The Thermofield dynamics~(TFD) state~\cite{Takahasi:1974zn,Takahashi:1996zn}, which is holographically dual to eternal AdS black hole~\cite{Maldacena:2001kr} in the context of AdS/CFT correspondence, can emerge in the low energy sector because of the emergent factorization of the Hilbert space. In $SU(N)$ case with a finite $N$, we consider a fine-tuned state in the form of
\begin{align}
	\Psi_\beta\,=\, \sum_{ \substack{\lambda\\ R=(\lambda,\lambda)\,,\, |\lambda|<N/2}} e^{-{\beta\over 4} E_\sun(R)} P(R, \{\phi\}) \ ,
\end{align}
where the Young tableau $R$ comprises two Young tableaux of identical shape $\lambda$, one of boxes and the other of anti-boxes (See Fig.~\ref{fig: emergent tfd state}). Hence these Young tableaux $R=(\lambda,\lambda)$ are self-conjugate, \textit{i.e.} $\overline{R}=R$. We limit the number of boxes in the Young tableau $\lambda$ to ensure that the Young tableau of boxes does not overlap with the Young tableau of anti-boxes. Such a large Young tableau will be irrelevant in the low energy description.

As the Hilbert space is factorized in the low energy sector, the state $\Psi_\beta$ becomes the TFD state with the inverse temperature $\beta$:
\begin{align}
	\Psi_\beta\,=\, \sum_{\lambda} e^{-{\beta\over 2} \big(E_\sun(\lambda) +{|\lambda|^2\over N}\big)} P(\lambda, \{\bar{\phi}\}) P(\lambda, \{\phi\})\ .\label{eq: tfd state schur}
\end{align}
In the strict large $N$ limit, it turns into the TFD state for the free loop gas, which is analogous to the \textit{generalized free field} in the holography:
\begin{align}
	\Psi_\beta\,=\,& \sum_{n=0}^\infty  e^{-{\beta N n\over 2}}\sum_{\nu \vdash n }  \prod_{j=1}^n{ \phi_j^{\nu_j}\bar{\phi}_j^{\nu_j} \over j^{\nu_j} \nu_j! }\ ,\cr
 \,=\,& \sum_{n=0}^\infty  e^{-{\beta N n\over 2}}\sum_{\nu \vdash n }  \prod_{j=1}^n{1\over  \nu_j! } (a_j^\dag)^{\nu_j}(\bar{a}_j^\dag)^{\nu_j}  \ .\label{eq: tfd state gff}
\end{align}
where $\phi_n$ and ${\partial\over \partial \phi_n}$ is replaced by $\sqrt{n} a^\dag_n$ and ${1\over \sqrt{n}}a_n$ respectively for the explicit Fock space representation~\cite{Jevicki:1991yi}. In the context of black hole, the collective field, $\bar{\phi}_n$, can be viewed as the mirror operator of $\phi_n$, arising from the smoothness condition of black hole horizon~\cite{Israel:1976ur}.

The construction of the mirror operator is state-dependent in the low energy description while it is state-independent at finite $N$ regime. The constraints in the matrix model give rise to the operator relation between the collective field $\phi_n$ and its mirror operator $\bar{\phi}_n$, which is independent of states:
\begin{align}
    P(\lambda,\bar{\phi})\,=\, P(\bar{\lambda},\phi)\ .
\end{align}
On the other hands, due to the absence of these constraints in the low energy description, the construction of the mirror operator becomes state-dependent:
\begin{align}
	\bar{\phi}_n\, \Psi_\beta\,=\, e^{{\beta  n\over 2} } n{\partial\over \partial \phi_n}\,  \Psi_\beta\ .
\end{align}
Here we considered the TFD state~\eqref{eq: tfd state gff} in the generalized free field limit. We also find that the state-dependent construction of the mirror operator with respect to the TFD state~\eqref{eq: tfd state schur} beyond the generalized free field limit can be approximated to be
\begin{align}
\label{eq:TFDmirrorop}
	P(\nu,\{\bar{\phi}_n\}) \,  \Psi_\beta\,\approx \, e^{{\beta\over 2} E_\sun(\nu) } P\bigg(\nu, \bigg\{ n {\partial \over \partial \phi_n}\bigg\}\bigg) \Psi_\beta \ .
\end{align}
The details are given in Appendix~\ref{app: state-dependent mirror operator}.

\subsection{Black Hole Complementarity}
\label{sec: black hole complementarity}

In the low energy description for the matrix model, accessible observables are limited because one cannot probe an observable associated with the large number of boxes. Therefore, the set of accessible observables in the singlet sector, which are polynomials of $\phi$'s with a finite degree, should have cutoff in the number of associated boxes. Due to the cutoff, the set of accessible observables denoted by $\cA^\trun_\phi$ does not form an algebra. This truncation by cutoff can also be observed in QFT on the curved background where a product of exponentially large number of matters would, in principle, produce a black hole operator, but we do not include the black hole operator in the set of accessible observables in the semi-classical calculations. This \textit{truncated algebra} $\cA^\trun_\phi$ has been explored in Refs.~\cite{Papadodimas:2012aq,Papadodimas:2013jku,Papadodimas:2013wnh,Papadodimas:2015jra,Banerjee:2016mhh,deBoer:2018ibj,DeBoer:2019yoe} to reinterpret the black hole complementarity~\cite{tHooft:1990fkf,Susskind:1993if} from the operator perspective.

The construction of the mirror operators $\bar{\phi}_n$ in the previous section elucidates a mechanism for the revised concept of black hole complementarity asserting that the degree of freedom in black hole interior are not independent of that of the outside~\cite{Papadodimas:2012aq,Papadodimas:2013jku,Papadodimas:2013wnh,Papadodimas:2015jra,Banerjee:2016mhh,deBoer:2018ibj,DeBoer:2019yoe}. At finite $N$, the mirror operator $\bar{\phi}_n$ is a polynomial of $\phi$'s associated with large Young tableaux. In the low energy sector, such large Young tableaux are beyond the truncated algebra $\cA^\trun_\phi$, allowing us to consider $\phi_n$ and its mirror operator $\bar{\phi}_m$ as independent degrees of freedom. As we begin to probe large Young tableaux (\textit{e.g.} higher-point correlations or the Hilbert space of the Hawking radiations after Page time), the constraints of the matrix model come into play. This leads to the breakdown of the semi-classical description breaks down.

The bulk locality is also an emergent phenomenon in the semi-classical description. $\bar{\phi}$ is viewed as an emergent quasiparticle that becomes independent in the low energy description, ensuring the exact bulk locality in this regime:
\begin{align}
    \bigg[\bar{\phi}_n,{\partial \over \partial \phi_m}\bigg]\,=\, 0 \ .
\end{align}
However, at finite $N$, the mirror operator $\bar{\phi}_n$ no longer maintains independence from $\phi$'s, resulting in the breakdown of the bulk locality:
\begin{align}
    \bigg[\bar{\phi}_1,{\partial \over \partial \phi_m}\bigg]\,=\,\bigg[P({\tiny \overline{\yng(1)}},\phi),{\partial \over \partial \phi_m}\bigg] \,\ne\, 0 \ .
\end{align}
This bulk non-locality, manifesting as an island in the black hole, will be explored in the next section.

\subsection{Page Curve, Island Conjecture and Holography of Information}
\label{sec: holography of information}

The emergent factorization in the matrix model can provide insight into the black hole information paradox~\cite{Hawking:1975vcx,Hawking:1976ra}. In the semi-classical description, introducing a cutoff to the accessible observables results in the factorization of the two truncated algebra, $\cA^\trun_\phi$ and $\cA^\trun_{\bar{\phi}}$, generated by $\phi_n$ and $\bar{\phi}_n$, respectively. In context of a black hole,  $\cA^\trun_\phi$ corresponds to observables outside the black hole, whereas $\cA^\trun_{\bar{\phi}}$ represents the observables inside the black hole that are difficult to construct by a finite product of the operators. During the black hole evaporation, the emitted Hawking radiation, which belongs to $\cA^\trun_\phi$, is entangled with the mirror modes in $\cA^\trun_{\bar{\phi}}$, leading to an increase in the entropy of the Hawking radiation. 

When the half of the black hole is evaporated, a point referred to as Page time, the number of the Hawking radiations become sufficiently large that the calculation of the entanglement entropy requires observables beyond the truncated algebra $\cA^\trun_\phi$. Adhering to the premise that the Hawking radiations and its mirror modes are independent, by overlooking the constraints in the matrix model even after the Page time, leads to a continuous increase in the entanglement entropy of the Hawking radiation past Page time.

\begin{figure}[t!]
\centering
\includegraphics[width=\linewidth]{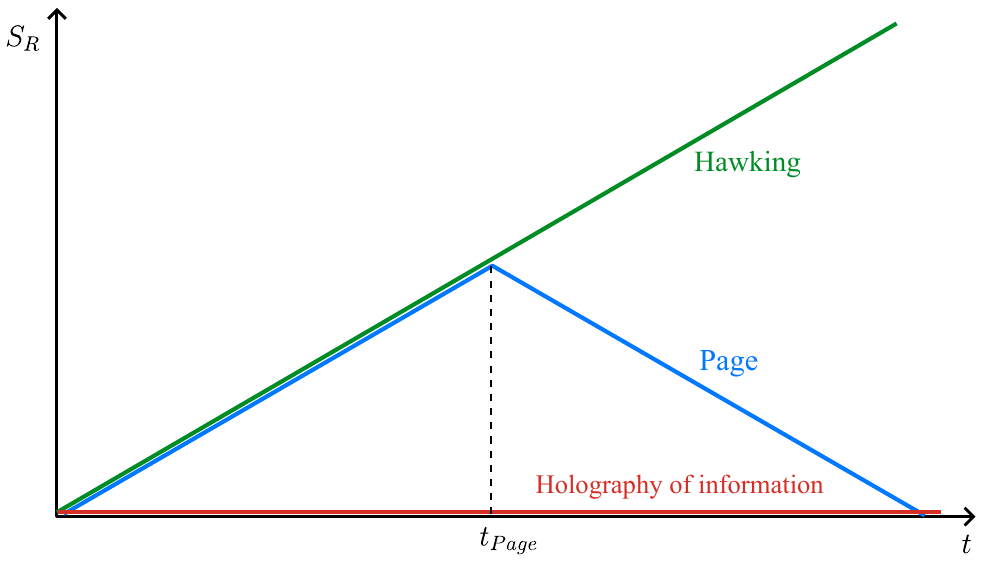}
	\caption{The entanglement entropy~(EE) of Hawking radiation. (green line) The EE keeps increasing as long as $\cA^\trun_\phi$ and $\cA^\trun_{\bar{\phi}}$ are regarded as independent. (blue line) After the Page time $t_{Page}$, $\cA^\trun_\phi$ and $\cA^\trun_{\bar{\phi}}$ are no longer independent, leading to the reduction in the EE. (red line) Considering the full algebra $\cA^{\text{\tiny full}}_{\phi}$ generated by $\phi$, rather than the truncated algebra $\cA^\trun_\phi$ of accessible observables, information about mirror operators is consistently present in $\cA^{\text{\tiny full}}_{\phi}$, bringing about the vanishing EE throughout the evaporation process. }
 \label{fig: ee of hawking radiation}
\end{figure}

After the Page time, the entanglement entropy calculations must incorporate the constraints in the matrix model, leading to the breakdown of the semi-classical description. Although we expect a decrease in the entanglement entropy following the Page curve~\cite{Page:1993df,Page:1993wv} (See Fig.~\ref{fig: ee of hawking radiation}), confirming this decline through methods beyond the semi-classical approaches, remains a significant challenge.

In spite of the breakdown of the semi-classical description after the Page time, the \textit{island} conjecture~\cite{Almheiri:2019psf,Almheiri:2019hni,Almheiri:2019yqk,Penington:2019kki,Almheiri:2019qdq,Almheiri:2020cfm,Bak:2020enw} has succeeded in reproducing the Page curve surprisingly by using semi-classical calculations. This conjecture proposes that the entanglement wedge of the Hawking radiation, referred to as an `island', emerges inside the black hole. The emergence of the island inside the black hole
, which extremizes the generalized entropy, plays a central role in the decrease of the entanglement entropy after the Page time. Nonetheless, the underlying reason why the semi-classical calculations by island managed to reproduce the Page curve has remained elusive.

 \begin{figure}[t!]
\centering
\includegraphics[width=\linewidth]{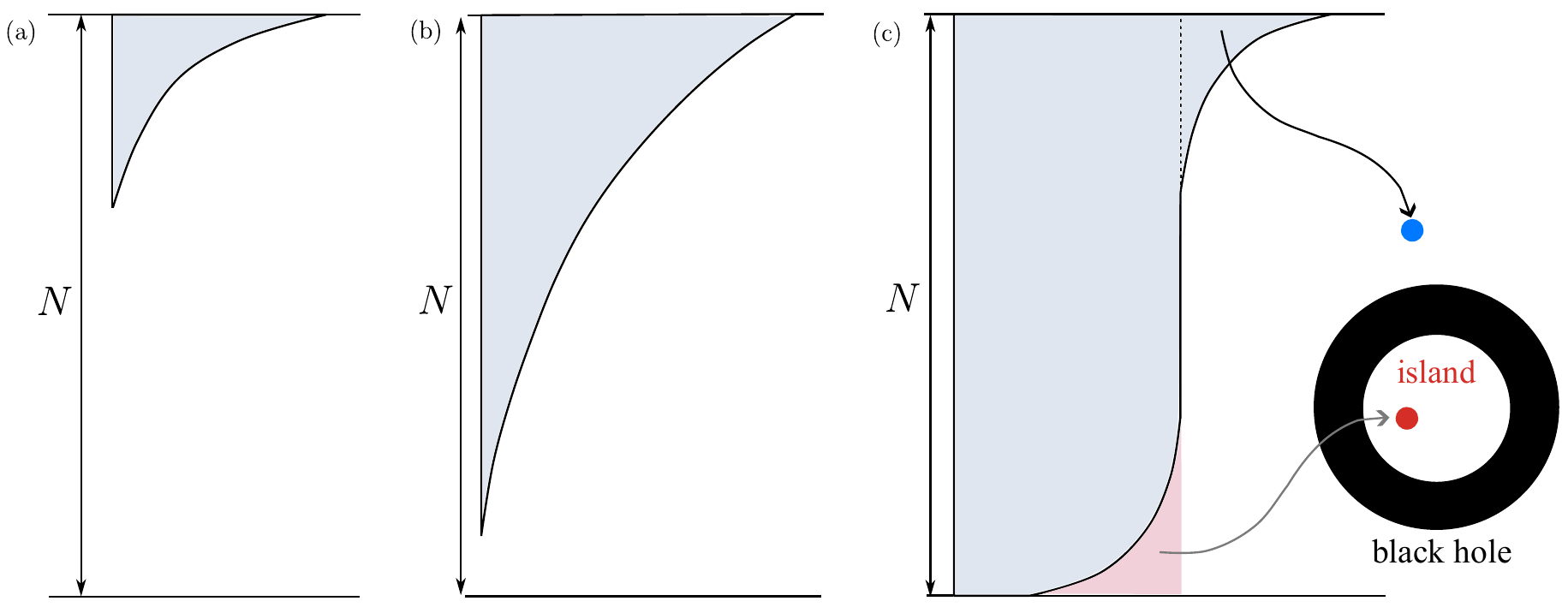}
	\caption{(a) Before the Page time, the entanglement entropy~(EE) calculation is involved with Young tableaux with finite numbers of boxes, in which the semi-classical description is valid. (b) After the Page time, the number of Hawking radiation becomes large enough to be involved with large Young tableaux. Generally, those large Young tableaux extend beyond the scope of the semi-classical description. (c) Certain large Young tableaux can effectively be represented as a pair of Young tableaux $(\lambda^\lyng,\lambda^\ryng)$, consisting of a finite number of anti-boxes and boxes, respectively. These Young tableaux can be dealt within the semi-classical framework, wherein the Young tableau $\lambda^\lyng$ (red) represents the collective excitation in the island inside the black hole while $\lambda^\ryng$ corresponds to the excitation outside the black hole.}
 \label{fig: island young tableau}
\end{figure}

The emergent factorization in the matrix model elucidates the origin of the island. After Page time, calculating the entanglement entropy of the Hawking radiation necessitates observables associated with large Young tableaux. Some of these Young tableaux can equivalently be viewed as a pair of Young tableaux $(\lambda^\lyng,\lambda^\ryng)$, consisting of a finite number of anti-boxes and boxes, respectively (See Fig.~\ref{fig: island young tableau}). As a result, they can be regarded as observables in $\cA^\trun_{\bar{\phi}} \otimes \cA^\trun_{\phi}$, enabling the semi-classical analysis. In other words, certain complicated operator can effectively be considered as two simple operators, one inside the black hole and the other outside. Hence, the entanglement wedge of the Hawking radiation includes the truncated algebra, $\cA^\trun_{\bar{\phi}} \otimes \cA^\trun_{\phi}$. The semi-classical analysis incorporating $\cA^\trun_{\bar{\phi}}$--- the island inside the black hole---can effectively capture the contributions of complicated operators.

Recently, the holography of information has been recently proposed, claiming that all information about the black hole interior is already available on the asymptotic boundary~\cite{Laddha:2020kvp,Chowdhury:2020hse,Raju:2020smc,Chowdhury:2021nxw,Raju:2021lwh,Chakraborty:2023los,Chakravarty:2023cll,deMelloKoch:2022sul}. Consequently, according to the holography of information, the entanglement entropy of the Hawking radiation would always be 0 throughout the black hole evaporation process.

The holography of information could be tantamount to asserting that the collective $\phi_n$'s can generate full operator algebra $\cA_\phi$ in the singlet sector of the matrix model which includes both $\cA^\trun_\phi$ and $\cA^\trun_{\bar{\phi}}$, \textit{i.e.} $\cA^\trun_\phi\cup \cA^\trun_{\bar{\phi}}\subseteq\cA^{\text{\tiny full}}_\phi$. Although one cannot access to the observables in the truncated algebra $\cA^\trun_{\bar{\phi}}$ by simple measurements by $\phi_n$'s, the information regarding the mirror operators $\bar{\phi}$'s is available in the operator algebra $\cA^{\text{\tiny full}}_\phi$ generated by $\phi$'s. Hence, disregarding the measurement, the entanglement entropy of the Hawking radiation remains at zero while the entanglement entropy based on the simple measurement will increase initially.

\section{Discussion}
\label{sec: discussion}

In this work, we have investigated the emergent factorization of the Hilbert space in the low energy description of the matrix models and its implications for understanding the black hole information paradox. By analyzing the collective description of a simple matrix model, we have demonstrated how the mirror operator---analogous to a hole in the band theory---emerges as a collective excitation in the low energy sector, thereby providing insights into the black hole complementarity. Furthermore, our study extends to discussing the emergence of TFD state from a fine-tuned state; elucidating the origin of the island that appears inside black holes; and offering a reinterpretation of the holography of information from the emergent factorization perspective.

Realizing the emergent factorization of the Hilbert space in realistic holographic models for black hole would be crucial. Although the matrix models examined in this work are simplistic enough to elucidate the mechanism behind the emergence of the collective mirror excitations, those models are, in essence, toy models that do not directly describe black holes. Therefore, the investigation on the emergence of the factorization in a concrete holographic matrix model, such as $\cN=4$ SYM, would be highly interesting. In particular, 
BPS black hole operator in large $N$ limit could provide a fruitful play ground to study the emergence of the mirror operator. Furthermore, it will be interesting to investigate whether the interior of black hole and the mirror operator emerges in the semi-classical description of fuzzball~\cite{Mathur:2005zp,Bena:2007kg}.

When the truncated ``algebra'' $\cA^\trun_\phi$ and $\cA^\trun_{\bar{\phi}}$ are assumed to be closed, the emergent factorization of Hilbert space in the low energy sector leads to type $III_1$ von Neumann factor. For example, the mirror observables in $\cA^\trun_{\bar{\phi}}$ can be viewed as the probe discussed in Ref.~\cite{Gesteau:2024dhj}, leading to the identification of the type $III_1$ factor. Note that we have showed that the mirror operators in $\cA^\trun_{\bar{\phi}}$ emerges naturally in the low energy sector, whereas Ref.~\cite{Gesteau:2024dhj} introduced the mirror operator by hand as a probe. However, the truncated ``algebra'' is not inherently closed, indicating that it should not be strictly defined as an algebra. Therefore, it becomes imperative to generalize the framework of von Neumann algebra and the Reeh–Schlieder theorem to accommodate a restricted set of accessible observables~\cite{Ghosh:2017gtw} in the low energy sector in a manner reminiscent of the same spirit as the Wilsonian renormalization group.

Another potential avenue for future work would be the ``non-factorization'' puzzle~\cite{Blommaert:2021fob,Saad:2021uzi,Choi:2022asl} in the context of AdS/CFT correspondence without ensemble average. The contribution of the Euclidean wormhole connecting two asymptotic AdS regions plays a crucial role in the random matrix behavior of the gravitational theories. Yet, conventionally holographically dual CFTs without the ensemble average would be factorized in the absence of ensemble averaging. Using the mechanism for the emergent factorization of the Hilbert space, one could consider a system where two CFTs are effectively factorized in the semi-classical description, yet remain non-factorized in their UV regime. In addition, the collective description of two decoupled CFTs could induce effective non-factorization~\cite{Jevicki:2015sla,Jevicki:2021ddf,Jevicki:2023yuh}, where the collective description for the diagonal $O(N)$ singlet sector of the two decoupled CFTs encompasses the non-singlet sector of each CFT. This diagonal $O(N)$ singlet sector of full system of two decoupled CFTs has more degrees of freedom compared to two $O(N)$ singlet sectors. These additional collective fields across the two CFTs could explain the non-factorization. In the context of the matrix quantum mechanics, the diagonal singlet sector can bring about the ensemble of non-singlet sectors associated with black holes~\cite{Kazakov:2000pm,Betzios:2021fnm,Betzios:2022pji,Betzios:2023obs}, giving rise to TFD-like state. From this perspective, the confinement/deconfinement transition may trigger a non-factorization/factorization transition, thereby offering further insight into this non-factorization puzzle.

\acknowledgements

We thank Seok Kim, Ki-seok Kim and Vinay Malvimat for discussion. J.Y. was supported by the National Research Foundation of Korea (NRF) grant funded by the Korean government (MSIT) (No.\ 2022R1A2C1003182, NRF-2023K2A9A1A01095488). J.Y. is supported by an appointment to the JRG Program at the APCTP through the Science and Technology Promotion Fund and Lottery Fund of the Korean Government. 
This is also supported by the Korean Local Governments - Gyeongsangbuk-do Province and Pohang City. This work is also supported by Korea Institute for Advanced Study (KIAS) grant
funded by the Korean government. A. J. was supported by the U.S. Department of Energy under contract DE-SC0010010. The work of D.M is supported by the \textit{Young Scientist Training Program} (YST) of the Asia Pacific Center of Theoretical Physics (APCTP).

\appendix

\section{Casimir for low-energy sector of $SU(N)$ matrix model}
\label{app: su(n) casimir}

In this appendix, we provide an explicit derivation of the decomposition of the energy eigenvalue $E_\sun(R)$~\eqref{eq: casimir su(n) modified} for a typical Young tableau in the low energy sector parametrized as $R \equiv (\lambda^{\lyng},\lambda^{\ryng})$ (See Fig.~\ref{fig: generic low energy young tableau}). The energy eigenvalue, $SU(N)$ Casimir, is given by 
\begin{align}
\label{eq:CasimirSUN}
E_{\text{\tiny SU(N)}}(R)  \,=\, \sum_{j=1}^N r_j \big(N +r_j -(2j-1)\big) -\frac{1}{N}\bigg(\sum_{j=1}^N r_j\bigg)^2\ .
\end{align}
where $r_j$ denotes the number of boxes in the $j$th row in the Young tableau $R$. The constituent tableaux $\lambda^\ryng$ and $\lambda^\lyng$ are parametrized by their corresponding row lengths $\lambda^\ryng \equiv (\lambda^\ryng_1,\lambda^\ryng_2,\dots,\lambda^\ryng_p)$ and $\lambda^\lyng \equiv (\lambda^\lyng_1,\lambda^\lyng_2,\dots,\lambda^\lyng_q)$ respectively. Thus, the parametrization of the full young tableau $R$ takes the form
\begin{equation}
r_i=
\begin{cases}
     \lambda^\lyng_1 +\lambda^\ryng_i \quad &\text{for} \quad 1\leq i \leq p\\
     \lambda^\lyng_1 \quad &\text{for} \quad p+1 \leq i \leq N-q\\
     \lambda^\lyng_1 -\lambda^\lyng_{N-i+1} \quad &\text{for} \quad N-q+1 \leq i \leq N
\end{cases}
\end{equation}
In terms of $\lambda^\lyng$ and $\lambda^\ryng$, the energy eigenvalue~\eqref{eq:CasimirSUN} can be expressed as
\begin{align}
    &E_{\text{\tiny SU(N)}}(R) \cr
    \,=\, & N\left(N\lambda^\lyng_1 +|\lambda^\ryng| -|\lambda^\lyng| \right)\cr
    &+N\lambda^\lyng_1(\lambda^\lyng_1 -N)+2\lambda^\lyng_1 (|\lambda^\ryng|-|\lambda^\lyng|)\cr
       &+\sum_{j=1}^p \lambda^\ryng_j(\lambda^\ryng_j-2j+1) +\sum_{k=1}^q \lambda^\lyng_k(\lambda^\lyng_k-2k+1)+2N|\lambda^\lyng|\cr
    &-\frac{1}{N}\left(N\lambda^\lyng_1 +|\lambda^\ryng| -|\lambda^\lyng| \right)^2\ ,
\end{align}
where $|\lambda^\lyng|$ and $|\lambda^\ryng|$ denotes the total number of boxes in $\lambda^\lyng$ and $\lambda^\ryng$, \textit{i.e.} $|\lambda^\lyng| \equiv\sum_{i=1}^p \lambda^\lyng_i$ and $|\lambda^\ryng|\equiv\sum_{i=1}^q \lambda^\ryng_i$. 
%
%
%
Finally, we obtain
\begin{equation}
\begin{aligned}
    E_{\text{\tiny SU(N)}}(R) &
    =  E_{\text{\tiny SU(N)}}(\lambda^\ryng)+  E_{\text{\tiny SU(N)}}(\lambda^\lyng)+\frac{2}{N}|\lambda^\lyng||\lambda^\ryng|\ ,
    \end{aligned}
\end{equation}
leading to the decomposition of the energy eigenvalue~\eqref{eq: casimir su(n) modified} of $R=(\lambda^\lyng,\lambda^\ryng)$.

\section{Casimir for semi-classical fluctuation around background $\Lambda^{(0)}$ in $U(N)$ matrix model}
\label{app: u(n) casimir background}

In this appendix, we derive the decomposition of the energy eigenvalue $E_\un (R)$ in Eq.~\eqref{eq: u(n) energy semi classical} for a Young tableau $R=(\lambda^\lyng,\lambda^\ryng)$. Recall that for the case of $U(N)$, the Casimir is given by
\begin{equation}
\label{eq:unCasimir}
    E_\un(R)=N \sum_{i=1}^N r_i + \sum_{i=1}^N r_i \big( r_i +1 -2 i \big)\ ,
\end{equation}
where $r_i$ denotes the number of boxes in the $j$th row of the Young tableau $R$. As in the previous appendix, the Young tableaux $\lambda^\lyng$ and $\lambda^\ryng$ are also parametrized by the number of boxes, $\lambda^\lyng_j$ and $\lambda^\ryng_j$.
\begin{align}
\label{eq:unparametrization}
	r_i\,=\, \begin{cases}
	\;\; M+\lambda^\ryng_i \quad & \text{for} \quad 1\leq i \leq p\\
	\;\; M \quad & \text{for} \quad p+1 \leq i \leq N-q\\
	\;\; M - \lambda^\lyng_{N-i+1} \quad & \text{for} \quad N-q+1 \leq i \leq N\\
	\end{cases}\ .
\end{align}
%
In the above, the row lengths of $\lambda^\ryng$ are given by $(r_1,r_2,\dots,r_p)$ while the row lengths of $\lambda^\lyng$ are given by $(s_1,s_{2},\dots,s_{q})$. Thus, plugging in the parametrization \eqref{eq:unparametrization} in \eqref{eq:unCasimir} and following an analogous computation as done in the preceding section, we get,
\begin{align}
	&E_\un(R)\cr
	\,=\,& N \big( \big|\lambda^\ryng\big| + \big| \lambda^\lyng \big|  \big) + NM^2 + 2M\big( \big|\lambda^\ryng\big| - \big| \lambda^\lyng \big|  \big) \cr
	&+ \sum_{i=1}^p \lambda^\ryng_i (\lambda^\ryng_i+1-2i)+ \sum_{i=1}^q  \lambda^\lyng_i(\lambda^\lyng_i  +1-2i)\cr
	\,=\,& NM^2 +2M\big( \big|\lambda^\ryng\big| - \big| \lambda^\lyng \big|  \big) + E_\un(\lambda^\ryng)\cr
 &+ E_\un(\lambda^\lyng)\ ,\cr
\end{align}
%

\section{Schur Polynomial identity for $\ast$-conjugated Young tableau}
\label{app: schur polynomial identity}

Let us consider a $N\times N$ unitary matrix $U$ of which eigenvalues are given by
\begin{align}
	U\,\sim\, \text{diag}(x_1,x_2,\cdots, x_N)\ .
\end{align}
The Schur polynomial corresponding to a Young tableau with $N$ boxes in one column is given by
\begin{align}
	P\big(\; \left.{\ytableausetup{smalltableaux,centertableaux} \tiny
	\begin{ytableau}
	 \\
	 \\
	\none[\tiny \cdot]\\
	\none[\tiny \cdot]\\
	\none[\tiny \cdot]\\
	 \\
	\end{ytableau} }\right\}{\scriptstyle N} \; ,U\big)\,=\, x_1x_2\cdots x_N\,=\, \det U\ .
\end{align}
Similarly, the Schur polynomial associated with the Young tableau with $N-1$ boxes in a single column can be written in terms of the determinant $\det U$ and the Schur polynomial of the inverse matrix $U^{-1}$ for one box:
\begin{align}
	P\big(\; \left. {\ytableausetup{smalltableaux,centertableaux} \tiny
	\begin{ytableau}
	 \\
	 \\
	\none[\tiny \cdot]\\
	\none[\tiny \cdot]\\
	\none[\tiny \cdot]\\
	 \\
	\end{ytableau} }\right\}{\scriptstyle N-1} \; ,U\big)=& \sum_{1\leqq i_1<i_2<\cdots< i_N\leqq N}x_{i_1}x_{i_2}\cdots x_{i_{N-1}}\cr
	&\hspace*{-2cm}=\ \det U \sum_{i=1}^N{1\over x_i}\,=\, \det U P_1\big(\; {\tiny \yng(1)}\;, U^{-1}\big)\ .
\end{align}
In general, a Young tableau composed of a single column with $N-n$ boxes can be identified with the Schur polynomial of $U^{-1}$ for $n$ boxes in a single column.
\begin{align}
	P_{N-n}\big(\; \left. {\ytableausetup{smalltableaux,centertableaux} \tiny
	\begin{ytableau}
	 \\
	 \\
	\none[\tiny \cdot]\\
	\none[\tiny \cdot]\\
	\none[\tiny \cdot]\\
	 \\
	\end{ytableau} }\right\}{\scriptstyle N-n} \; ,U\big)&= \sum_{1\leqq i_1<i_2<\cdots< i_{N-n}\leqq N}x_{i_1}x_{i_2}\cdots x_{i_{N-n}}\cr
	&\hspace*{-1cm}=\det U \sum_{1\leqq i_1<i_2<\cdots< i_n\leqq N }{1\over x_{i_1}x_{i_2}\cdots x_{i_n}}\cr
    &\hspace*{-1cm}=\det U P_{n}\big(\; \left. {\ytableausetup{smalltableaux,centertableaux} \tiny
	\begin{ytableau}
	 \\
	 \\
	\none[\tiny \cdot]\\
	\none[\tiny \cdot]\\
	\none[\tiny \cdot]\\
	 \\
	\end{ytableau} }\right\}{\scriptstyle n} \; ,U^{-1}\big)\ .
\end{align}
For a generic Young tableau $\lambda$, the corresponding Schur polynomial can be written as
\begin{align}
	&P(\lambda,U)\,=\, {\det \left(x_j^{\lambda_i+N-i}\right) \over {\displaystyle \prod_{1\leqq i<j\leqq N}}\! (x_i-x_j) }\cr
	\,&= {1\over {\displaystyle \prod_{1\leqq i<j\leqq N}\ } (x_i-x_j)} \begin{vmatrix}
	x_1^{\lambda_1+N-1} & x_2^{\lambda_1+N-1} & \cdots & x_N^{\lambda_1+N-1} \\
	x_1^{\lambda_2+N-2} & x_2^{\lambda_2+N-2} & \cdots & x_N^{\lambda_2+N-2}\\
	\vdots & \vdots & \ddots & \vdots\\
	x_1^{\lambda_N} & x_2^{\lambda_N} & \cdots & x_N^{\lambda_N}\\
	\end{vmatrix}\ ,\cr
 &\label{eq:genSchur}
\end{align}
where $\lambda_j$ denotes the number of boxes in the $j$th row of the Young tableau $\lambda$.

Denoting the number of boxes in the $j$th row of the $\ast$-conjugated Young tableau $\lambda^\ast$ by $\nu_j\equiv (\lambda^\ast)_j$, we have  
\begin{align}
	\lambda_i\,=\, M- \nu_{N-i+1}\ ,
\end{align}
where we took the $\ast$-conjugation~\eqref{eq:starconjugation} with respect to a $N\times M$ background Young tableau $\Lambda^{(0)}$.
%
%
%
%
%
From Eq.~\eqref{eq:genSchur}, we obtain the relation between two Schur polynomials $P(\lambda,U)$ and $P(\lambda^\ast,U^{-1})$:
\begin{align}
    &P(\lambda,U)\,\nonumber\\
    &=\ {{\displaystyle \prod_i} x_i^{M} \over {\displaystyle \prod_{1\leqq i<j\leqq N}\ } (x_i^{-1}-x_j^{-1})} 
 \begin{vmatrix}
	x_1^{-\nu_1-N+1} & x_2^{-\nu_1-N+1} & \cdots & x_N^{-\nu_1-N+1}\\
	\vdots & \vdots & \ddots & \vdots\\
	x_1^{-\nu_{N-1}+1} & x_2^{-\nu_{N-1}+1} & \cdots & x_N^{-\nu_{N-1}+1}\\
	x_1^{-\nu_N} & x_2^{-\nu_N} & \cdots & x_N^{-\nu_N} \\
	\end{vmatrix}\nonumber\\
	&=\ \big[\det U\big]^M P(\lambda^\ast , U^{-1})\ .
\end{align}
This provides the relation between an arbitrary Young tableau and its corresponding $\ast$-\textit{conjugated} one. 

\section{State-dependent Construction of Mirror Operator}
\label{app: state-dependent mirror operator}

In this appendix, we present the derivation of the state-dependent construction of mirror operators in Eq.~\eqref{eq:TFDmirrorop}. The TFD state $\Psi_{\beta}$~\eqref{eq: tfd state schur}, which emerges in the low energy sector from a fine-tuned state $\Psi_\beta$, is given by
\begin{align}
	\Psi_\beta\,=\, \sum_{\lambda} e^{-{\beta\over 2} \big(E_\sun(\lambda) +{|\lambda|^2\over N}\big)} P(\lambda, \{\bar{\phi}\}) P(\lambda, \{\phi\})\ ,
\end{align}
where $\lambda$ runs over Young tableaux with a finite number of boxes. The action a Schur polynomial of the mirror operator $\bar{\phi}_n$ corresponding to a $\cO(N^0)$ Young tableau $\nu$ on the TFD state can be evaluated to be
\begin{align}
	&P(\nu,\{\bar{\phi}_n\})\,\Psi_\beta\cr
 \,=\,& \sum_{\lambda} e^{-{\beta\over 2} \big(E_\sun(\lambda) +{|\lambda|^2\over N}\big)} P(\nu,\{\bar{\phi}_n\})\,P(\lambda, \{\bar{\phi}\}) P(\lambda, \{\phi\})\ ,\cr
    \,=\,& \sum_{\lambda,\kappa} e^{-{\beta\over 2} \big(E_\sun(\lambda) +{|\lambda|^2\over N}\big)} {c_{\nu \lambda}}^{\kappa}P(\kappa,\{\bar{\phi}_n\}) P(\lambda, \{\phi\})\ ,
\end{align}
where ${c_{\nu \lambda}}^{\kappa}$ denotes the tensor product coefficients. Since both the Young tableau $\lambda$ and $\nu$ have $\cO(N^0)$ numbers of boxes, the coefficient ${c_{\nu \lambda}}^{\kappa}$ does not vanish only when the Young tableau $\kappa$ in the summation satisfies the condition
\begin{align}
    |\kappa|\,=\,|\lambda|+|\nu|\ .\label{eq: non-vanishing condition}
\end{align}
Recall $\bar{\phi}_n$ corresponds to the mirror creation operator. Hence, let us now consider the action of the annihilation operator on the TFD state. Noting the relation $\phi_n$ with $\sqrt{n} a^\dag_n$ and ${\partial\over \partial \phi_n}$ with ${1\over \sqrt{n}}a_n$, we will evaluate the action of $P\bigg(\nu, \bigg\{ n {\partial \over \partial \phi_n}\bigg\}\bigg)$ corresponding to Young tableau $\nu$ of $\cO(N^0)$ boxes on the TFD state:
\begin{align}
    &P\bigg(\nu, \bigg\{ n {\partial \over \partial \phi_n}\bigg\}\bigg)\Psi_\beta\cr
    \,=\, &\sum_{\kappa} e^{-{\beta\over 2} \big(E_\sun(\kappa) +{|\kappa|^2\over N}\big)}P(\kappa, \{\bar{\phi}\}) \cr
    &\qquad\times P\bigg(\nu, \bigg\{ n {\partial \over \partial \phi_n}\bigg\}\bigg)P(\kappa, \{\phi\})\ ,\label{eq: action of annihilation on tfe}
\end{align}
where $\kappa$ runs over the Young tableau of $\cO(N^0)$ boxes. To evaluate $P\bigg(\nu, \bigg\{ n {\partial \over \partial \phi_n}\bigg\}\bigg)P(\kappa, \{\phi\})$, we will utilize the product of two Schur polynomials $P(\nu, \{ \bar{\phi} \})$ and $P(\kappa, \{\phi\})$ at finite $N$. At this finite $N$, $\phi_n$ and $\bar{\phi}_n$ are not independent, leading to a linear combination of Schur polynomials of $\phi$'s as the outcome. Among these resultant Schur polynomials, those that have finite number of Young tableaux correspond to the result of $P\bigg(\nu, \bigg\{ n {\partial \over \partial \phi_n}\bigg\}\bigg)P(\kappa, \{\phi\})$. 
\begin{align}
   &P\bigg(\nu, \bigg\{ n {\partial \over \partial \phi_n}\bigg\}\bigg)P(\kappa, \{\phi\})\cr
   \,=\,& P(\nu, \{ \bar{\phi} \})P(\kappa, \{\phi\})\big|_{\text{finite number of boxes}}\ .
\end{align}
Since the Young tableau $\nu$ of anti-boxes is equivalent to the conjugate Young tableau $\bar{\nu}$ of boxes, one may also use the product of $ P(\bar{\nu}, \{ \phi \})$ and $P(\kappa, \{\phi\})$. Then, using the tensor product of the Young tableau, the product can be decomposed into a linear combination of Schur polynomials. 
\begin{align}
   &P\bigg(\nu, \bigg\{ n {\partial \over \partial \phi_n}\bigg\}\bigg)P(\kappa, \{\phi\})\cr
   \,=\,& P(\bar{\nu}, \{ \phi \})P(\kappa, \{\phi\})\big|_{\text{finite number of boxes}}\ ,\\
   \,=\,& \sum_{\lambda} {c_{\bar{\nu}\kappa}}^\lambda P(\lambda, \{ \phi \})\big|_{\lambda\,:\, \text{finite number of boxes}}\ ,
\end{align}
where we also select Young tableau $\lambda$ that has a finite number of boxes. Therefore, one can write the right-hand side of Eq.~\eqref{eq: action of annihilation on tfe} as follows.
\begin{align}
    &P\bigg(\nu, \bigg\{ n {\partial \over \partial \phi_n}\bigg\}\bigg)\Psi_\beta\\
    \,=\, &\sum_{\kappa,\lambda} e^{-{\beta\over 2} \big(E_\sun(\kappa) +{|\kappa|^2\over N}\big)} {c_{\bar{\nu}\kappa}}^{\lambda}P(\kappa, \{\bar{\phi}\})P(\lambda, \{\phi\})\ ,
\end{align}
where $\kappa$ and $\lambda$ run over Young tableaux with finite numbers of boxes. Using the identities of the tensor-product coefficient
\begin{align}
    {c_{\bar{\nu}\kappa}}^{\lambda}\,=\, {c_{\bar{\nu}\bar{\lambda} }}^{\bar{\kappa}}\,=\, {c_{\nu \lambda}}^{\kappa}\ ,
\end{align}
we finally have
\begin{align}
    &P\bigg(\nu, \bigg\{ n {\partial \over \partial \phi_n}\bigg\}\bigg)\Psi_\beta\cr
    \,=\, &\sum_{\kappa,\lambda} e^{-{\beta\over 2} \big(E_\sun(\kappa) +{|\kappa|^2\over N}\big)} {c_{\nu\lambda}}^{\kappa}P(\kappa, \{\bar{\phi}\})P(\lambda, \{\phi\})\ .
\end{align}
For comparison, we can rewrite $P(\nu,\{\bar{\phi}_n\})\,\Psi_\beta$ as
\begin{align}
	&P(\nu,\{\bar{\phi}_n\})\,\Psi_\beta\cr
    \,=\,& \sum_{\lambda,\kappa} e^{-{\beta\over 2} \big(E_\sun(\lambda) +{|\lambda|^2\over N}\big)} {c_{\nu \lambda}}^{\kappa}P(\kappa,\{\bar{\phi}_n\}) P(\lambda, \{\phi\})\\
    \,=\, &e^{{\beta\over 2} \big(E_\sun(\nu) +{|\nu|^2\over N}\big)}\sum_{\lambda,\kappa}  e^{-{\beta\over 2} \big(E_\sun(\kappa)+ \Delta E(\nu,\lambda,\kappa) +{|\kappa|^2\over N}\big)} \cr
    &\hspace{25mm}\times {c_{\nu \lambda}}^{\kappa}P(\kappa,\{\bar{\phi}_n\}) P(\lambda, \{\phi\})\ ,
\end{align}
Here, we define $\Delta E(\nu,\lambda, \kappa)$ by
\begin{align}
    \Delta E(\nu,\lambda,\kappa)\,\equiv\, & E_\sun(\nu)+ E_\sun(\lambda) -E_\sun(\kappa)  \cr
    &+{1\over N}\big(|\nu|^2+|\lambda|^2 - |\kappa|^2 \big)\\
    \sim\,&\cO(N^0)\ ,
\end{align}
where we can confirm that it is of order $\cO(N^0)$ due to the non-vanishing condition~\eqref{eq: non-vanishing condition} for the tensor-product coefficient. Consequently, if we ignore $\Delta E(\nu,\lambda,\kappa)\sim \cO(N^0)$ in the exponent, we can find the state-dependent construction of the mirror operator:
\begin{align}
    P(\nu,\{\bar{\phi}_n\})\,\Psi_\beta\,\approx\, e^{{\beta\over 2} \big(E_\sun(\nu) \big)} P\bigg(\nu, \bigg\{ n {\partial \over \partial \phi_n}\bigg\}\bigg)\Psi_\beta
\end{align}
Strictly speaking, neglecting terms of order $\cO(N^0)$ in the exponent would be equivalent to the free loop gas limit in Section~\ref{sec: emergent tfd state} where one can immediately obtain the state-dependent construction.

\bibliography{matrix}

\end{document}